\documentclass[12pt]{elsarticle}

\usepackage{amssymb}
\usepackage{amsthm}
\usepackage{booktabs}
\usepackage{tabularray}
 \usepackage{amsmath}
\usepackage{longtable}
\usepackage{pdflscape}
\usepackage{adjustbox}
\usepackage{pifont}
\usepackage{graphicx}
\usepackage{natbib}

\journal{ }

\begin{document}

\begin{frontmatter}

\title{Protecting Africa's Future: Cybersecurity Strategies for Child Safety, Learning, and Skill Acquisition in Tanzania}
\date{ }

\author[1,*]{Ezekia Gilliard}
\author[4]{Abdul Maziko}
\author[1]{Gideon Rwechungura}
\author[6]{Ahmed Abubakar Aliyu}
\author[3]{Erasto Kayumbe}

\address[1]{Mwalimu Nyerere University of Agriculture and Technology, Mara, Tanzania}
\address[4]{Tanzania Public Service College, Tabora, Tanzania}
\address[3]{Tanzania Atomic Energy Commission, Arusha, Tanzania}
%\address[2]{Wuhan University, Hubei, China}
%\address[5]{Tsinghua University, Beijing, China}
\address[6]{Kaduna State University  Kaduna 800283 Nigeria }
\address[*]{Corresponding author: giezekias@gmail.com}

\begin{abstract}
Today, children across Africa are at a growing risk from the Internet. Dangers include harmful content, violence, exploitation, abuse, and neglect. All these have increased due to increased mobile and Internet technology use, which not only places their safety at risk but also affects their ability to learn essential skills for their future.
This paper provides an overview of the unique challenges faced by third-world African countries in ensuring the online safety of children while also supporting their developmental needs. It highlights effective practices and policies adopted by other nations to safeguard children from online threats and enhance their digital literacy. We are focusing on sharing the best practices and policies other countries have used to protect children from abuse and help them succeed in the digital world. The study emphasizes the online safety strategies, legal frameworks, and recommendations specific to the United Republic of Tanzania, along with the significance of international collaborations with organizations like UNICEF and the UN. The goal is to provide African policymakers, educators, and cybersecurity professionals with practical guidance and recommendations to strengthen child online safety initiatives both within and beyond the continent.
\end{abstract}

\begin{keyword}
Digital Skills, CyberSpace, cybersecurity, child safety, online learning, internet safety
\end{keyword}

\end{frontmatter}

\section{Introduction} % 

Development in digital technologies has shifted many life aspects, including education in Africa. The increasing digital innovations present an opportunity to better their education, vocational training, and employment prospects for the disadvantaged. Even as this change came with its benefits, it was faced with various cybersecurity threats that placed children at risk. These threats include bullying, scamming, pornography, and many more online threats. This means that the eruption of COVID-19 pandemic has accelerated digitization and increased usage of the internet for remote learning, thus increasing the risks for children. Most African countries do not have in place cybersecurity awareness and literacy programs and policies for the protection of children \cite{reid2016children,adedoyin2023covid}.

In the last seven years, the continent's internet accessibility has expanded twofold, propelled by the swift construction of fiber optic infrastructure, now three times as extensive as it was seven years prior \cite{wirajing2023revisiting}. The growing use of mobile phones is responsible for a significant portion of this increase in online activity. This progress has imbued millions of African children and youth with new digital opportunities but also exposed them to cybersecurity threats.

The Internet and the digital world provide children with unprecedented educational and learning opportunities. Children can use the Internet to carry out learning, social and creative activities. As emphasized in the United Nations Committee on the Rights of the Child's General Comment No. 25 on the Rights of the Child in relation to the digital environment, the Internet has great potential in realizing children's rights \cite{child_general_2021}. However, accessing digital platforms also carries certain risks, including cyberbullying, harassment,child pornography, gambling, and online scams, this new aspect of life have huge impacts on the well-being and pursuit of children toward digital learning. Like most African countries, Tanzania has a few strategies and policies to ensure child safety from these cyber threats.  Therefore, due to the rapid growth of cybercrimes, quick intervention is called for to safeguard and uphold the right of children to learn and develop in a safe digital environment. Research suggests that the increasing adoption of digital technology exposes Tanzanian adolescents to cyber risks \cite{onditi2018tanzanian}. 

Interpol and UNICEF funded the "Disrupting Harm in Tanzania" report, which reveals that 97\% of children aged 12 to 17 access the internet from home rather than at school or other locations. This finding highlights the significant responsibility of parents and caregivers, as most children use smartphones to go online, often without supervision. At least 2\% of kids reported taking nude photos or videos of themselves in the previous year. The same percentage of kids admitted to letting someone else snap a picture of them naked \cite{dcruz_launch_2022}.

\begin{figure}
    \centering
    \includegraphics[width=0.7\linewidth]{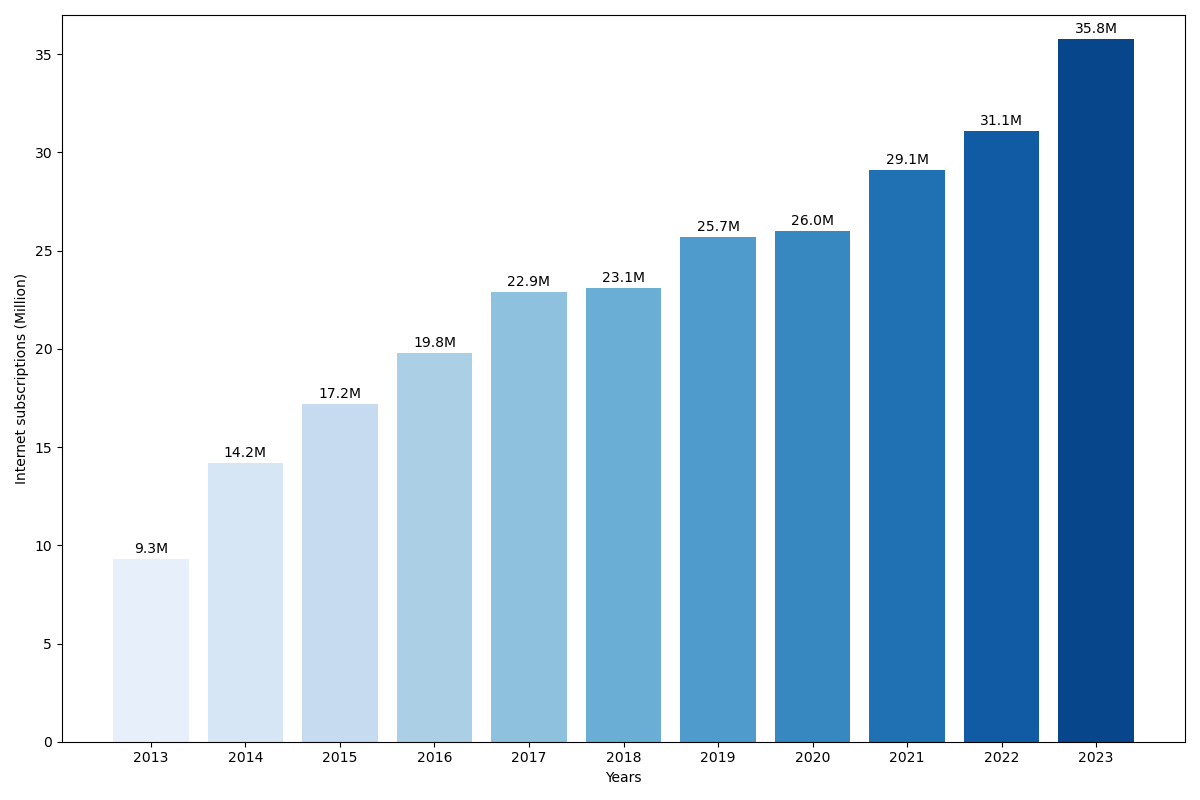}
    \caption{Over the past decade, internet subscriptions have seen nearly a fourfold increase, rising from 9.3 million in 2013 to 35.8 million in 2023. (Source: TCRA Quarterly Reports, 2023)}
    \label{fig:enter-label}
\end{figure} 

Tanzania has implemented ICT policies to transform into an information-driven society, but there are significant concerns regarding cyber threats to children in education. A study found high rates of cell phone and internet use among secondary school students, with many experiencing online violence, cyberbullying, and sexual exploitation \cite{onditi2018tanzanian}. Another study revealed that 42\% of students reported cyberbullying others, while 58\% experienced cybervictimization \cite{shapka2019adolescent}. Despite these challenges, Tanzania faces obstacles in implementing effective cybersecurity measures, including limited digital literacy, inadequate resources, and infrastructure issues \cite{kayumbecybersecurity} . For example, commercial banks in Tanzania have implemented cybercrime mitigation strategies, but attacks still occur, highlighting the need for stronger systems \cite{mwita2023assessing}. These findings underscore the importance of developing culturally appropriate education and intervention programs that involve global, regional, and national partners to address cyber threats and protect Tanzanian children in the rapidly evolving digital landscape.

\begin{table}
\centering
\caption{Risk for Children in the Digital Environment: revised typology of risks (OECD, 2021)}
\label{tab:child_risks}
\resizebox{\linewidth}{!}{%
\begin{tabular}{llccc} 
\toprule
\textbf{Risk Categories} & \textbf{Content Risks} & \textbf{Conduct Risks} & \textbf{Contact Risks} & \textbf{Consumer Risks} \\ 
\midrule
\textbf{Cross-cutting Risks*} & \multicolumn{4}{c}{Privacy Risks (Interpersonal, Institutional  Commercial)} \\
 & \multicolumn{4}{c}{Advanced Technology Risks (e.g. AI, IoT, Predictive Analytics, Biometrics)} \\
 & \multicolumn{4}{c}{Risks on Health  Wellbeing} \\ 
\hline\hline
\textbf{Risk Manifestations} & Hateful Content & Hateful Behaviors & Hateful Encounters & Marketing Risks \\
 & Harmful Content & Harmful Behaviors & Harmful Encounters & Commercial Risks \\
 & Illegal Content & Illegal Behaviors & Illegal Encounters & Financial Risks \\
 & Disinformation & User-generated Problematic Behaviors & Other Problematic Encounters & Security Risks \\
\bottomrule
\end{tabular}
}
\end{table}

As the member of the "We Protect Together" global alliance and International Telecommunication Union (ITU), Tanzanian government has tried to combat the growing challenges that come with these technological advances. Some of these initiatives include the ICT policy for basic education of 2007, the Cybercrime Act No. 13 of 2015, the Child Act R.E 2019, Zanzibar’s Children’s Act and numerous institutions to ensure the safety of cyberspace. However, research indicates that many people, including public servants in Tanzania, have limited awareness of cybercrimes and cyberlaws, which hinders the effective implementation of these law. The role of parents and teachers in protecting children from online harm is crucial. They provide essential guidance and supervision to help ensure children's safety while they navigate the internet \cite{puspita2021role,ilmanto2021problems}. However, most teachers and parents lack the critical knowledge and skills to effectively monitor and guide children online.

In this research paper, the focus is on analyzing the specific cybersecurity challenges that children in Tanzania are currently encountering also provides thorough recommendations. These suggestions can help in implementing effective strategies and policies to enhance child safety, facilitate better learning experiences, and promote skill acquisition. These recommendations are derived from studying and incorporating best practices observed in other countries, ensuring a comprehensive and well-informed approach to addressing the cybersecurity needs of Tanzanian children.

%https://www.tcra.go.tz/tcra-news/cybersecurity-and-safety-key-for-development
%chrome-extension://efaidnbmnnnibpcajpcglclefindmkaj/https://www.itu.int/en/ITU-D/Cybersecurity/Documents/COP/20-00802_COP-Policy_Brief.pdf
\section{Analysis of Current Cybersecurity Strategies in Tanzania}
Online safety is important for children, especially when it comes to ensuring that they study online without distraction and are sure that they are safe. This section looks at how well internet safety strategies in Tanzania protect children. We review current policies and practices, assess their effectiveness, and identify areas that need attention to strengthen the protection of children in the digital world. Understanding these factors allows us to identify areas for improvement and recommend improvements to protect young consumers.

\subsection{Existing Policies and Procedures}
Tanzania has made notable strides in addressing internet security challenges, yet several issues remain. A primary obstacle is the limited awareness among internet users regarding the risks associated with its misuse. The rise in smartphone usage has increased these challenges, raising significant concerns about children's online safety. This is due to the fact that most children access the internet using smartphones as opposed to other devices. As depicted in Figure \ref{fig:activities}, children primarily use the internet for watching videos and social media, which can significantly compromise their safety. These concerns also discourage children from using digital devices for educational purposes because they present cybersecurity threats. The government has taken decisive action to bolster cyberspace resilience. It has enacted legislation, established institutions, and devised policies. Here are a few examples of these efforts:

\begin{figure}
    \centering
    \includegraphics[width=1\linewidth]{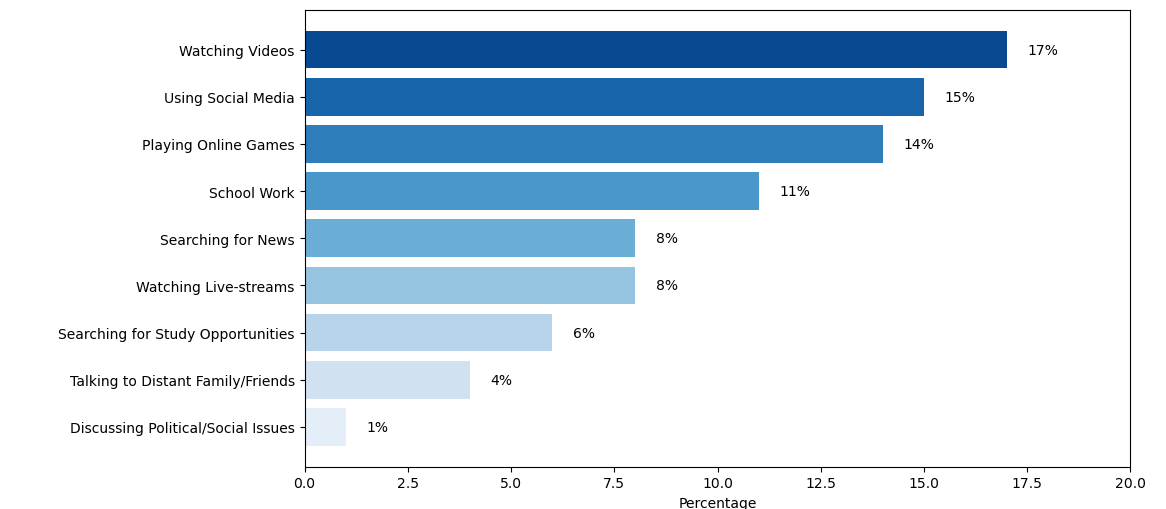}
    \caption{Distribution of Online Activities Among Children (Data source: Livingstone, S., Kardefelt Winther, D., \& Saeed, M. (2019). Global Kids Online Comparative Report. Innocenti Research Report. Florence: UNICEF Office of Research – Innocenti.}
    \label{fig:activities}
\end{figure}

\begin{enumerate}
    \item [\ding{109}] The Tanzania Communications Authority (TCRA) \footnote{https://www.tcra.go.tz/}, established in 2003, occupies a pivotal position in this domain. Tanzania entrusts the TCRA with communications and broadcasting regulation, as well as cybersecurity oversight. Since its establishment, TCRA has championed digital literacy, IT infrastructure development, and substantial achievement in cybersecurity. Through the Tanzania Computer Emergency Response Team (TZ-CERT) \footnote{https://www.tzcert.go.tz/} , TCRA has made efforts to identify, raise, and support the security skills and culture of various groups. In collaboration with other government institutions, TZ-CERT has established Child Online Protection (COP) by giving out advice and tips to parents and guardians on online safety for children. With these efforts, Tanzania was ranked second in Africa by the International Telecommunication Union (ITU) in the Global Cybersecurity Index Report 2020 (GCI)\footnote{https://www.itu.int/en/ITU-D/Cybersecurity/Pages/global-cybersecurity-index.aspx}. TCRA and other related bodies enforced legal and regulatory measures, which contributed to Tanzania's exemplary performance.
    
    Although the TCRA has performed well in many areas, it has not focused much on child protection. Most of its guidelines focus on general internet safety, with limited directives specifically aimed at protecting children online. Authorities have primarily focused on communication channel safety, but they have not explicitly targeted the online risks faced by children.

    \item [\ding{109}] The foundation of Tanzania's ICT structure is the National IT Policy, which was issued as early as 2003, paving the way for its revision, which was finally done in June of 2016. This initiative has caused many innovations, such as sector liberalization and the implementation of the fiber optic National Information Technology Backbone (NICTBB) for interconnecting regions as well as districts across the country. This has in turn resulted in reduced internet costs and increased access to ICT services. For example, Tanzania now offers the most affordable internet bundles in East Africa, similar to those in Nigeria \footnote{https://www.statista.com/chart/29144/cost-of-mobile-data-in-africa/}.

    However, challenges remain, particularly concerning last-mile connectivity and educational access in rural areas. Limited infrastructure, high costs, and unreliable internet access contribute to a persistent digital divide \cite{MUSTAFA2024102380,brewer2022last}. These issues hinder equitable learning opportunities and impede the development of digital literacy among students and faculty in rural higher education settings. While NICTBB has been instrumental in bridging some of the digital divides, these areas continue to struggle with connectivity issues. Additionally, although the policy takes a comprehensive approach to ICT development and security, it lacks specific measures to safeguard children online. The general provisions for internet safety are in place, but there is a noticeable gap when it comes to implementing targeted strategies that protect children in the digital landscape. This oversight highlights the need for a more focused approach to ensure the safety of the younger population in an increasingly connected world.

    \item [\ding{109}] The Cyber Crimes Act, 2015, is one of Tanzania's other reliable pieces of legislation, as it has also helped to reduce cybercrimes on the internet. It establishes a legal framework to address various forms of cybercrime, including unauthorized access to computers and information systems, online fraud, child pornography, etc. Studies have found that it has increased awareness among users about proper internet conduct and the potential consequences of misuse \cite{mambile2020cybercrimes}. The legislation aims to protect users in the digital space, addressing challenges posed by the evolving nature of technology and online interactions.
    
    However, recent research indicates that cybercrime rates remain high in Tanzania, despite efforts by the police force. Also, despite the Act's prohibition of child pornography, cyber harassment, and exploitation, among other offenses, there remains a need for further implementation and enforcement \cite{massawe2023preventing}. The word 'child' appeared just nine times throughout the entire act, indicating a clear need for laws and policies that are more tailored to child protection in the digital age. These specific laws aim to appropriately advance and enforce specific safeguards for children within the context of digital platforms. Currently, the Tanzanian Parliament is in the process of collecting stakeholder comments on the Child Protection Laws (Miscellaneous Amendments) Bill.

    \item [\ding{109}] The Tanzania Police Cyber Crime Unit is a special unit within the Tanzania Police Force, created to deal with the country's growing cybercrime threat. The unit is responsible for investigating various cyber-related crimes, including online fraud, hacking, identity theft, and cyberbullying. Along with its responsibilities for investigations, this unit collaborates with other government agencies, such as TCRA, and international organizations to increase its capacity to deal with complex cyber threats. Even though this unit has made significant contributions to the implementation of the Cyber Crime Act 2015, the task of tracking cyber criminals and bringing them to justice remains challenging in the absence of trained personnel and modern tools. However, while the unit is important in enforcing general cybercrime laws, its efforts to focus on the protection of children online are still ongoing.

    \item [\ding{109}] The National Online Child Protection System is an emerging initiative within Tanzania's broader efforts to protect children in the digital environment. Recognizing the unique risks that children face online, such as cyberbullying, exploitation, and exposure to inappropriate content, the framework aims to take a comprehensive approach to children's online safety. The Ministry of Social Development, Gender, Women, and Special Needs, in collaboration with the Tanzania Communications Authority (TCRA) under the Tanzania Computer Emergency Response Team (TZ-CERT), are working together on a campaign on child protection online. Their goal is to raise awareness and protect children from online dangers while enabling them to take advantage of digital opportunities. The system educates children, parents, teachers, and the community about their responsibilities in protecting kids online and provides resources to deal with cybercrime.

    \item [\ding{109}]The National Child Online Protection Framework aims to develop guidelines and best practices to ensure safe online use for children, addressing areas such as content regulation, privacy protection, and enforcement of online abuse reporting mechanisms. The system emphasizes education and awareness, implementing programs to equip children, parents, and educators with the knowledge and tools needed to navigate the digital landscape safely. Additionally, the framework advocates strong collaboration between government agencies, educational institutions, non-governmental organizations, and technology companies to effectively utilize resources and expertise. However, despite its ambitious goals, the system is still in the initial stages of development and has not yet reached full implementation across the country. To effectively protect children in an increasingly digital world, continuous improvement of its guidelines, effective implementation, and the adoption of more children are essential.

    \item [\ding{109}] Education programs in Tanzania play an important role in promoting online safety and digital literacy among children, parents, and educators. These programs are essential in equipping people with the necessary skills and knowledge to navigate the digital world responsibly. Digital literacy initiatives, often led by NGOs, educational institutions, and international organizations, form the basis of these efforts. These programs aim to teach children safe internet practices, such as identifying online threats, understanding privacy settings, and reporting incidents of cyberbullying. At the same time, public awareness campaigns inform parents and guardians about the potential risks associated with their children's online activities. These campaigns usually include workshops, seminars, and informational materials aimed at helping parents monitor their children's online behavior.

    In 2007, Tanzania launched its ICT Policy for Primary Education to transform the country into an information-driven society and enhance education. The policy aimed to integrate ICT into primary, secondary, and vocational education, ensuring that children and youth have the essential skills and precautions needed when they enter cyberspace. Nonetheless, its implementation has faced challenges, including inadequate teachers training and insufficient IT equipment in schools. Despite these challenges, some students have acquired basic ICT skills, especially in using the Internet and software applications. International communities and NGOs have supported various ICT projects in Tanzania, contributing to the government's Education Sector Development Plan by donating ICT equipment such as computers and building computer labs.

    \item [\ding{109}]The Child Act, R.E. 2019, and Zanzibar’s Children’s Act 2011 contain important provisions regarding child sexual abuse material (CSAM) and the use of children in pornographic performances. The Law of the Child Act, R.E. 2019, is applicable in Mainland Tanzania, while the Children’s Act 2011 is applicable in Zanzibar. It's important to note that Zanzibar's Children's Act 2011 is the only law in Tanzania that criminalizes possessing and gaining access to CSAM. However, the legislation does not explicitly address the livestreaming of child sexual abuse, and there is no specific provision that criminalizes this practice. This is very important because there is an increase in these events on social media. Additionally, there are loopholes in the law that result from the absence of provisions prohibiting online grooming for sexual purposes and sexual extortion committed in the online environment. The Child Act R.E. 2019 is currently undergoing amendments.
\end{enumerate}

However, these efforts have primarily focused on protecting government and business systems, often ignoring issues related to the online safety of children and young users.

\begin{figure}
    \centering
    \includegraphics[width=1\linewidth]{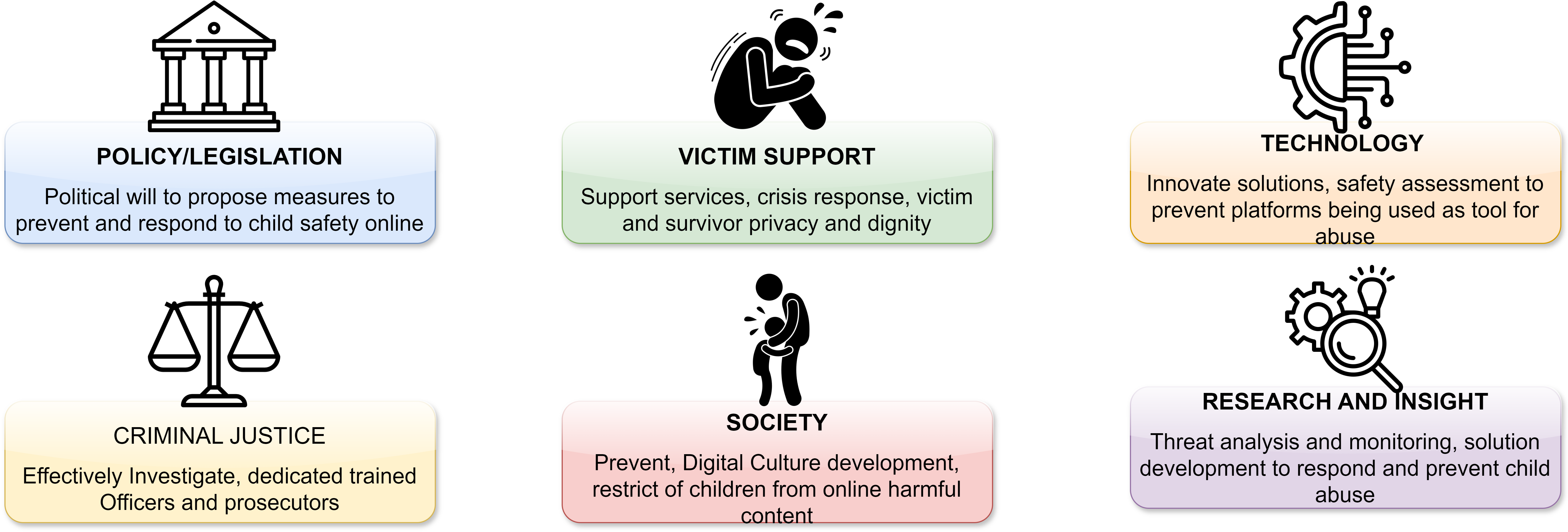}
    \caption{We Protect Global Strategic Pillars for Child Online Safety: The diagram illustrates six key pillars—Policy/Legislation, Victim Support, Technology, Criminal Justice, Society, and Research and Insight—that form a comprehensive strategy for protecting children online.}
    \label{fig:weprotect}
\end{figure}

\section{Recommendations for Tanzania}
Encouraging children to use the internet safely is crucial in order to enable them to acquire valuable skills and knowledge without any difficulties. All relevant stakeholders must be involved to overcome online child protection obstacles. Although that Tanzania has made significant strides in cybersecurity by ranking second in Africa by the ITU's Global Cybersecurity Index, but still requires stronger laws and regulations pertaining to child online safety. This indicates that policies that prioritize keeping children safe online are desperately needed.
Including important individuals in the development and implementation of cybersecurity plans that meet Tanzania's demands is of utmost importance. These important individuals include NGOs, international organizations, IT corporations, parents, teachers, and the government. By increasing awareness and strengthening their capacity to respond, the International Telecommunication Union assists nations like Tanzania in addressing global cybersecurity challenges. The WeProtect Global Alliance collaborates with several organizations to end the sexual exploitation of children online. In addition, UNICEF is crucial in defending children's rights and formulating laws that support and shield kids from harm when they use technology.

Tanzania has much to gain by examining the approaches taken by other nations in handling similar issues, such as China, the USA, Australia, the UK, the EU, Canada, and Japan. These nations have robust legal frameworks, policies, and initiatives that may serve as a model for Tanzania in terms of ensuring children's online safety. The suggestions section outlines specific steps that the government, parents, educators, tech firms, non-governmental groups, and international organizations may take to improve children's online safety. These measures encompass educating the public, adhering to cybersecurity regulations, endorsing worthwhile initiatives, and assisting children in acquiring secure internet skills. These suggestions aim to provide Tanzanian children with a safer online environment in which they can develop and study while remaining secure.

\begin{table}[htbp]
\centering
\caption{Comparative Analysis of Child Online Safety Legislation and Initiatives}
\begin{tabular}{@{}p{2cm}p{6cm}p{6cm}@{}}
\toprule
\textbf{Country} & \textbf{Key Legislation / Initiative} & \textbf{Relevance to Tanzania} \\ \midrule
China            & Law of the People's Republic of China on Protection of Minors \footnote{$http://en.npc.gov.cn.cdurl.cn/2020-10/17/c_674690.htm$}, Cybersecurity Law 2017 \footnote{https://digichina.stanford.edu/work/translation-cybersecurity-law-of-the-peoples-republic-of-china-effective-june-1-2017/}        & Ensuring strict content regulation and parental controls              \\ 
USA              & Children’s Online Privacy Protection Act 1998  (COPPA) \footnote{https://www.ftc.gov/legal-library/browse/rules/childrens-online-privacy-protection-rule-coppa}  & Protecting children’s data and ensuring parental involvement           \\ 
Australia        & The Online Safety Act, 2021 \footnote{$https://www.aph.gov.au/Parliamentary_Business/Bills_LEGislation/Bills_Search_Results/Result?bId=r6680$}                             & Integrating digital literacy into the school curriculum                \\ 
UK               & UK Online Safety Act, 2023 \footnote{https://www.legislation.gov.uk/ukpga/2023/50}                       & Designing online platforms with child safety as a priority             \\ 
European Union   & General Data Protection Regulation (GDPR) 2016 \footnote{https://gdpr-info.eu/}, Better Internet for Kids (BIK+) \footnote{https://eur-lex.europa.eu/legal-content/EN/ALL/?uri=CELEX:52012DC0196} & Comprehensive data protection and child online safety tools \\ 
Canada           & Digital Charter Implementation Act, 2022 \footnote{https://www.parl.ca/legisinfo/en/bill/44-1/c-27}, MediaSmarts \footnote{https://mediasmarts.ca/}                      & Promoting digital literacy and safe online environments                \\ 
Japan            & Act on Development of an Environment that Provides Safe and Secure Internet Use for Young People 2009 & Multi-stakeholder approach to child online safety  \\ \bottomrule
\end{tabular}
\label{tab:comparative_analysis}
\end{table}

\subsection{Legislation and Governance}
To make the internet safer for children in Tanzania and help them learn useful skills, it's important to learn from successful strategies used in other countries. By studying and adjusting these methods, Tanzania can create a strong system to protect kids online.

One of those countries is China, where the Law of the People's Republic of China on Protection of Minors \footnote{$http://en.npc.gov.cn.cdurl.cn/2020-10/17/c_674690.htm$} mandates parental controls and age checks on online sites, making the internet safer for youngsters. This is exactly what Tanzania needs, as the law enforces strict regulations, mandates child mode on platforms, and regulates gaming. Furthermore, these platforms necessitate online real-name registration to enforce these restrictions. A To keep youngsters safe, the Cybersecurity Law \footnote{https://digichina.stanford.edu/work/translation-cybersecurity-law-of-the-peoples-republic-of-china-effective-june-1-2017/} protects data and controls content. Additionally, the law discusses the regulation of critical information infrastructure, which may host or process large amounts of data about minors with stricter cybersecurity requirements. Penalties for non-compliance are severe, which include fines, suspension of licenses, and even criminal charges. These laws have the potential to strengthen Tanzania's legal framework and establish a safer online environment for children, facilitating the acquisition of new skills.

The Children's Online Privacy Protection Act 1998 (COPPA) \footnote{https://www.ftc.gov/legal-library/browse/rules/childrens-online-privacy-protection-rule-coppa} in the US controls the acquisition of personal data from children under 13. COPPA compels online services to acquire parental consent before collecting data, making the internet safer for kids. Initiatives like Common Sense Media assist parents and teachers in creating safe online learning environments by providing them with the tools they need to make the internet safer for children. The eSafety Commissioner \footnote{$https://www.aph.gov.au/Parliamentary_Business/Bills_LEGislation/Bills_Search_Results/Result?bId=r6680$} in Australia emphasizes instruction and robust problem reporting. Proactively addressing online safety, the eSafety Commissioner offers educational programs, parent resources, and a cyberbullying reporting system. Teaching digital literacy in classrooms helps kids be safe online and develop critical skills. New cyber abuse offenses targeting harassment and image-based abuse (revenge porn) also allow the eSafety Commission to order the removal of intimate images shared without consent, particularly when minors are involved. These incidents are very common in developing countries and sometimes lead to children committing suicide.

The UK's Age-Appropriate Design Code (Children's Code)\footnote{https://www.legislation.gov.uk/ukpga/2023/50} sets rigorous requirements for children's internet services. This rule requires platforms to use clear, safe methods to protect children's privacy and data. Tanzania might learn from the Code's focus on kid-safe digital activities. The European Union’s General Data Protection Regulation (GDPR) 2016 \footnote{https://gdpr-info.eu/} includes requirements for protecting children's data. Along with Better Internet for Kids, which provides information to children, parents, and teachers, the GDPR takes a comprehensive approach to internet safety. These guidelines safeguard data and educate users about internet threats. Additionally, it grants individuals the right to erasure, also known as the right to be forgotten, especially in cases where data was collected during their childhood and they no longer consent to its processing. In Japan, the government, commercial sector, and civil society collaborate to provide a secure online area. The Act on Development of an Environment that Provides Safe and Secure Internet Use for Young People encourages education and filtering software to safeguard minors from hazardous information. This cooperative concept explains how several groups can protect children online.

The criminal justice system lacks the necessary tools to enforce all these laws, rendering them worthless. We recommend collaboration and new technology for investigations. It's also critical to educate judges and police about internet child sexual exploitation in order to improve enforcement. The main takeaway is that Tanzania's government must establish more precise legislation to prohibit online child exploitation, enhance child abuse cases, and meet international standards. New regulations might mandate parental approval and data protection, like COPPA. The Tanzania Communications Regulatory Authority (TCRA) should oversee online platform safety measures, including age checks and parental controls. The TCRA might enforce data protection and internet safety rules. The Ministry of Education and child welfare officials should spearhead internet safety and digital learning activities, as well as establish necessary tools. Reporting issues and promoting digital literacy in schools will help kids and parents use the internet securely. Tanzania can protect children online and help them learn valuable skills.

\subsection{Support for victims}
Specialized treatment is necessary to guarantee the recovery and wellbeing of victims of child sexual exploitation and abuse on the internet. This will help children develop empathy and understanding as they get older and learn to recognize and attend to the needs of others. To offer complete therapy and reduce the danger of retraumatization, a robust counseling service and support system must strictly adhere to the present child protection framework.

One effective tactic is the establishment of one-stop rescue centers, which have the ability to promptly provide resources in cases of abuse. These institutions would house social services, mental health counseling, medical care, and legal assistance, among other services. This integrated approach streamlines the response to abuse while also lessening the need for survivors to continuously retell their experiences in a variety of venues, thereby preventing future harm to them.By guaranteeing that kids who encounter online threats or harassment have access to prompt and efficient support, the Weprotect Framework(Model National Response see Figure \ref{fig:weprotect}) offers a means of assisting victims and fostering a sense of safety and empowerment in them. In order to enhance the effectiveness of these centers, relevant organizations need to actively support the creation of one-stop shops for gathering evidence and conducting rescue operations across various areas. 

The UK's Child Exploitation and Online Protection Command (CEOP)\footnote{https://www.ceop.police.uk/Safety-Centre/}, as a component of the national crime agency, provides tools to help parents and children report online abuse. Comparable to that is the Stop Child Abuse (SCA) \footnote{https://www.nationalcac.org/stop-child-abuse-and-neglect-scan/} Campaign in the United States, which includes a number of agencies, including the FBI, that conduct awareness campaigns and offer resources for reporting and preventing child abuse online. Tanzania has the potential to use these models to create analogous policies and processes for these establishments, ensuring uniformity in mistreatment management.

Schools, doctors, court social workers, and other child welfare stakeholders also desperately need funding for capacity building. By strengthening the education and experience of these professionals, we can improve the quality of support and care provided to victims of child sexual exploitation and abuse on the internet. This training should emphasize a multidisciplinary approach and foster collaboration among professionals in the sectors of health, education, law enforcement, and social services in order to effectively meet the complex needs of victims.
We recommend building a long-term structure for one-stop evidence collection and rescue operations to ensure the viability of these activities. We should incorporate regular protocol evaluations, continual professional growth, and adjustments to reflect new developments and industry best practices into this system. We can ensure that victims receive the care and justice they require while also constructing a robust support system that can adjust to the ever-changing terrain of child abuse and exploitation on the internet.

\subsection{Education}
Education is essential for tackling the complex problem of online child sexual exploitation and abuse, particularly when the goal is to provide a secure environment that enables children to learn new skills. By providing children, adolescents, their caregivers, instructors, and the general public with essential knowledge, we may greatly improve awareness and comprehension of these crucial matters.

Education functions as a potent instrument for providing knowledge and authority to individuals, allowing youngsters to securely navigate online environments where they may acquire understanding and enhance their abilities. In a developing nation like Tanzania, cyberthreats and all crimes related to the internet are a new thing. Educating the community and children on internet usage best practices could increase awareness, instill values, and foster digital citizenship. These programs should prioritize raising awareness among children and adolescents, who are the main users of digital platforms, as well as their caregivers, educators, and the general public. By enhancing knowledge through education, we may establish a more knowledgeable society that not only safeguards children but also facilitates their acquisition of skills in a secure online setting.

Integrating cybersecurity education into the national curriculum is crucial \cite{ahmad2021cybersecurity}. Integrating cybersecurity into digital literacy lessons has advanced this sector in several nations. Courses cover healthy relationships, violence prevention, and bullying prevention. By including these themes in the curriculum, we ensure that children learn about online safety from an early age and have the tools to safely navigate the digital world while learning new skills. Social advocacy should include wide public awareness beyond formal schooling. Media and IT companies should be involved in these efforts to improve the message and reach more people. By participating in these campaigns, media and technology companies may influence public opinion, awareness, and online education safety for children. The media can raise awareness of the risks by showcasing representative examples and advocating for solutions while protecting privacy. Additionally, children must actively participate in these programs \cite{shillair2022cybersecurity,nicholson2021training}.

Encouragement to share ideas for internet security can provide more efficient and child-friendly results. By helping kids create a safe online environment, we empower them to boldly explore, learn, and grow. Internet safety is essential for youngsters to gain new digital skills. We must teach safety, privacy, and internet responsibility. An antivirus, firewall, and spam filter are essential for protecting a child's learning environment from cyberattacks \cite{alshabibi2021cybersecurity}. Maintaining a safe learning environment requires ongoing security improvements, especially for contactless devices and network infrastructure. Internet safety education includes teaching kids when to share personal information, the risks of exposing their location, and the importance of privacy. 

Furthermore, this instruction equips caregivers and educators with the knowledge and skills to assist children in using digital platforms in a secure manner. For parents with limited digital literacy, particularly those with basic reading and writing abilities, an understanding of the fundamental principles of social media usage is crucial. This encompasses the administration of public and private accounts, the processing of friend requests, and the imposition of social media penalties. It is imperative that we are able to manage the technical threats that are posed to us, such as viruses, fraud, worms, and phishing. It is imperative that users are aware of the measures they can take to avoid malware, spyware, and hacking risks. Furthermore, it is imperative to address the issues of cyberbullying, online grooming, and the potential risks associated with the disclosure of excessive personal information. Such dangers may impede children from studying online in a secure manner. The protection of privacy is of paramount importance to the security of the Internet. For example, the Canadian Digital Charter establishes guidelines for online safety and learning, while initiatives such as MediaSmarts provide educational resources on digital literacy and internet safety for young people. This facilitates comprehension of the potential consequences of disclosing personal information online, the enduring nature of digital actions, the significance of safeguarding digital assets, and the risks associated with identity theft. The provision of this information to users will facilitate the creation of a safer digital environment in which children can learn and develop. Education is an effective means of equipping children with the skills they require to navigate the digital landscape safely. The integration of cybersecurity and internet safety into national education and the involvement of multiple stakeholders in advocacy initiatives can collectively contribute to the development of a safer and more informed society \cite{sauglam2023systematic}. This comprehensive approach ensures the safety of children while facilitating their ability to navigate and enhance their online experiences with assurance.

\subsection{Stakeholder and the industry engagement}
It is of the utmost importance for these organizations to promptly submit any illicit or harmful information to the relevant authorities, emphasizing the necessity for collaboration across different departments to effectively address these issues. Ensuring a secure environment for children is a collective responsibility, with each individual bearing accountability to one another, the government, and parents. It is imperative that each sector fulfill its duties in order to collectively assist children in having a safer online space. It is critical to include stakeholders and the information and communication technology (ICT) sector in order to ensure the online safety of children and facilitate their learning and acquisition of skills. The ICT sector has a pivotal role in establishing efficient systems and methods to handle illicit information that may endanger children. 
An important opportunity for the ICT sector is to utilize developing information technologies, such as artificial intelligence (AI), to identify and report instances of child sexual exploitation and abuse. Artificial intelligence can play a crucial role in detecting trends and behaviors that could suggest exploitation, allowing for faster action \cite{ramesh2024ai,singh2024role}. Research emphasizes that we must not only employ technology as a temporary solution. Instead, a comprehensive child protection policy should incorporate it, carefully considering the need for child safety and safeguarding private rights. This method ensures that technology is used in accordance with child safety measures and upholds children's rights.
In order to achieve this objective, it is crucial for technology developers in the sector to undergo thorough training on child rights, child protection, and the special difficulties associated with online sexual exploitation and abuse. The goal of this workshop is to enhance developers' understanding of the ethical issues and obligations associated with designing technology that engages children. Developers may design tools and systems that effectively identify and prevent exploitation while also respecting children's rights \cite{faraz2022child}.
The ICT sector should also embrace the "safety by design" approach, which places a high priority on safeguarding and promoting the well-being of children right from the beginning of technology development. The concept of "maximizing the welfare of children" should guide this model, ensuring that all technical solutions prioritize the safety and well-being of young users. Furthermore, it is critical for ICT businesses to actively listen to and incorporate the viewpoints of young people in the development and execution of these technologies. By employing this approach, they may provide solutions that are both user-friendly and highly successful, aligning closely with the unique experiences and requirements of children.
Furthermore, in Tanzania, industry participation necessitates specialized measures such as content filtering and age verification, which are critical components alongside larger obligations. Technology businesses operating within the nation should adopt more stringent content filtering measures, such as those used in countries such as Japan and the United Kingdom. These procedures serve to screen and eliminate improper information that has the potential to cause harm to children, thereby ensuring a more secure online environment.
As a critical step, all technology enterprises should be required to implement age verification measures. These methods serve to mitigate the availability of inappropriate content for kids, therefore diminishing the likelihood of their exposure to hazardous material. Mandating companies to promptly notify local authorities of any instances of child exploitation is imperative. This not only ensures the pursuit of appropriate legal measures but also contributes to a broader effort to protect children in the digital sphere.
Essentially, stakeholder and ICT sector involvement is critical in establishing a more secure online environment for children in Tanzania. Through the establishment of efficient systems for managing content, responsible utilization of emerging technologies, and the implementation of strong practices for moderating information and verifying age, the industry can actively contribute to protecting children while also facilitating their learning and the acquisition of new abilities. The combined efforts of these individuals and organisations, driven by a commitment to safeguarding children and the principles of designing for safety, will enhance the digital experience for children by making it more secure and empowering.

\begin{longtable}{p{0.20\textwidth} p{0.28\textwidth} p{0.34\textwidth} p{0.18\textwidth}}
\caption{Roles of Stakeholders in Ensuring Child Online Safety} \\
\toprule
\textbf{Stakeholder} & \textbf{Role} & \textbf{Responsibilities} & \textbf{Impact} \\
\midrule
\endfirsthead

\toprule
\textbf{Stakeholder} & \textbf{Role} & \textbf{Responsibilities} & \textbf{Impact} \\
\midrule
\endhead

\midrule
\multicolumn{4}{r}{\textit{Continued on next page}} \\
\bottomrule
\endfoot

\bottomrule
\endlastfoot

Government & Policy and Regulation & Develop, enforce, and monitor legal frameworks to protect children online. & High  \\
\midrule

Educators & Digital Literacy and Awareness & Integrate digital literacy programs into school curricula and train educators. & High  \\
\midrule

NGOs & Advocacy, Support, and Training & Provide resources, training, and continuous support to children, parents, and educators. & Medium  \\
\midrule

International Organizations & International Policy Alignment, Funding, and Capacity Building & Partner with local bodies to implement best practices, provide financial support, and align policies with global standards. & High  \\
\midrule

Parents and Guardians & Supervision, Guidance, and Awareness & Monitor children’s online activities, set rules for internet use, and educate them on risks. & Medium  \\
\midrule

\end{longtable}

\subsection{Research and Publications}
Education, international organizations, universities, and research institutions play a crucial role in addressing the challenging issues associated with ensuring the online safety of children, facilitating their learning, and fostering their skill development. These groups play a crucial role in doing research, collecting information, and developing innovative strategies to safeguard children from online hazards while facilitating their development in the digital realm.

The process of systematically collecting and evaluating data is of utmost importance. The "National Response Model" emphasizes the importance of doing a thorough analysis of the situation to assess the effectiveness of measures aimed at safeguarding children on the internet. An exemplary instance is the annual publication of the "National Research Report on Internet Use of Minors" by the Central Committee of the Communist Youth League in China \footnote{https://www.gqt.org.cn/}. This study provides a comprehensive data analysis, monitoring the patterns and vulnerabilities associated with children's internet usage on an annual basis. Reports of this nature are critical for identifying emerging hazards, such as sexual solicitation and exploitation, as well as providing guidance for policy development and the implementation of measures to mitigate these risks.

Tanzania should implement comparable measures to methodically gather and evaluate data about children's internet usage and the hazards they encounter. Local governments have the option to employ research methodologies utilized by international organizations like UNICEF, such as the "Global Children Online" \footnote{http://globalkidsonline.net/} and Disrupting Harm" surveys \footnote{https://www.end-violence.org/disrupting-harm}. These approaches provide insights into children's online interactions and potential risks. We can customize them to fit Tanzania's specific circumstances. Tanzania can develop a comprehensive awareness of the unique issues its children encounter on the internet.

International entities such as UNICEF, the International Telecommunication Union (ITU), and WeProtect Global Alliance significantly contribute to the support of these research endeavors. In addition to offering research methodologies, they also provide venues for sharing findings and best practices across other nations. The establishment of this worldwide partnership is critical for developing a comprehensive understanding of internet safety concerns for children and devising synchronized approaches to address them.
Universities and research institutes are significant contributors to this endeavor. They contribute to the advancement of novel tools, technologies, and tactics aimed at enhancing children's online safety by conducting comprehensive investigations and generating research based on empirical data. Universities can spearhead research initiatives to evaluate the efficacy of various pedagogical approaches, assess the outcomes of digital literacy initiatives, or advance the creation of artificial intelligence technologies for identifying online hazards. Research endeavors should prioritize not only identifying dangers but also discovering novel remedies for widespread implementation.
Furthermore, it is imperative for research institutes to maintain ongoing communication with industry specialists to ensure the implementation of their results. This collaboration can result in the development of innovative technologies and systems that prioritize the well-being and safety of children. Research has the potential to result in improved age verification systems, more intelligent content filtering tools, and educational initiatives that instruct youngsters on safe online navigation.

Universities and research institutions should actively engage in international exchanges and partnerships in addition to doing research. Individuals can actively contribute to a worldwide endeavor to enhance online safety for children by disseminating their study to a global audience and assimilating knowledge from other nations. This collaborative approach ensures the integration of research into a global effort to protect children in the digital era, rather than its isolation.

In conclusion, it can be stated that there is a significant responsibility placed upon those entities involved in the field of education, international organisations, universities, and research institutions to ensure the safety of children when using the internet. By conducting rigorous research, employing systematic data gathering methods, and fostering international collaboration, these organizations may develop and disseminate innovative approaches to address the increasing complexities of the digital realm. It is of paramount importance that their work enhances online safety for young people while also fostering their future-oriented skill development.

\begin{table}[]
\caption{Summary of Online Child Protection Tools and Technologies}
\label{tab:my-table}
\resizebox{\textwidth}{!}{%
\begin{tabular}{@{}ll@{}}
\toprule
\textbf{Tool/Technology} &
  \textbf{Key Features and Relevance to Tanzania} \\ \midrule
Parental Control Software &
  \begin{tabular}[c]{@{}l@{}}The software is capable of setting time limits and\\ blocking specific websites and settings.\\ \\ If utilized correctly, it can assist in the monitoring\\ and restriction of access. Notable examples include\\ Net Nanny and Qustodia.\end{tabular} \\ \midrule
Content Filtering Tools &
  \begin{tabular}[c]{@{}l@{}}This tool blocks access to harmful content by \\ filtering inappropriate material and it has real updates.\\ Examples are CleanBrowsing and OpenDNS.\end{tabular} \\ \midrule
Monitoring Software &
  \begin{tabular}[c]{@{}l@{}}This tool lets parents, teachers, or guardians keep an \\ eye on social media activity and get alerts for anything \\ suspicious. Some examples are Bark and Family Time.\end{tabular} \\ \midrule
Safe Browsing Tools &
  \begin{tabular}[c]{@{}l@{}}These tools provide child-safe search engines and \\ automatically block adult content, ensuring that \\ children access age-appropriate materials only.\\ Examples are Kiddle and Safe Search kids.\end{tabular} \\ \midrule
Age-Appropriate Social Networks &
  \begin{tabular}[c]{@{}l@{}}These platforms are designed for children, featuring \\ age-appropriate content and parental oversight \\ to protect children from harmful online interactions.\\ Kidzworld and Kidzvuz are some of the examples.\end{tabular} \\ \midrule
Digital Literacy Platforms &
  \begin{tabular}[c]{@{}l@{}}These platforms offer lessons on digital citizenship \\ and interactive learning modules to teach children \\ how to navigate the internet safely.\\ Some of the organization include Common Sense\\ media and Childnet\end{tabular} \\ \midrule
Mobile Safety Apps &
  \begin{tabular}[c]{@{}l@{}}These apps allow parents to control app usage, \\ track locations, and monitor calls and messages, \\ offering a comprehensive mobile safety solution.\\ Examples are Life360 and MMGurdian.\end{tabular} \\ \midrule
AI-Based Threat Detection &
  \begin{tabular}[c]{@{}l@{}}These tools use AI to detect cyberbullying and \\ provide real-time alerts for risky online behavior, \\ offering proactive threat management.\\ Examples are GoGuardian and Securly.\end{tabular} \\ \midrule
Safe Gaming Platforms &
  \begin{tabular}[c]{@{}l@{}}These platforms feature moderated chat functions \\ and offer age-appropriate games, ensuring a safer \\ gaming environment for children.\end{tabular} \\ \midrule
Government-Backed Tools &
  \begin{tabular}[c]{@{}l@{}}These tools are part of national initiatives or \\ public-private partnerships designed to protect \\ children online, offering a robust framework \\ supported by state policies\end{tabular} \\ \midrule
School-Based Monitoring Systems &
  \begin{tabular}[c]{@{}l@{}}These systems track students' online activity \\ during school hours and ensure safe browsing, \\ offering a controlled and secure learning environment.\\ Tool like DansGuardian, Smoothwall, LightSpeed etc.\end{tabular} \\ \bottomrule
\end{tabular}%
}
\end{table}

\section{Conclusion}
Our research underscores the critical need for a comprehensive, coordinated approach to online child protection in Tanzania. Addressing the challenges identified requires collaboration among government bodies, the ICT industry, social organizations, and caregivers. Together, these stakeholders can create a secure digital environment that supports children's learning, skill development, and overall well-being.

Lessons from countries such as China, the USA, the UK, Australia, and Japan highlight the importance of implementing robust, child-specific online protection laws. These nations have established strong legal frameworks that effectively address the unique risks children face online, from exploitation to cyberbullying. For Tanzania, adopting similar legislation would be a crucial step in safeguarding children and ensuring a safer online experience.

The rise of online child sexual exploitation and abuse presents a growing concern, with these issues becoming more complex and widespread. However, increased awareness across society and a stronger commitment to action offer hope. This collective consciousness and coordinated response indicate that current challenges can be effectively tackled.

The "National Response Model" proposed by the WeProtect Global Alliance offers valuable guidance for developing and executing national strategies to combat these issues. Countries like China, with the world's largest population of young internet users, have proactively raised public awareness and strengthened legal protections. Japan, too, has made significant strides in leveraging technological innovation to prevent online exploitation. Both nations, however, acknowledge the need for ongoing efforts in research, data collection, and enhancing response mechanisms.

The ICT industry plays a crucial role in detecting and reporting online child exploitation and abuse. The development of technology-based solutions, stricter regulations, and accountability systems has led to significant progress. Yet, it remains essential to balance the protection of children's rights with ensuring their safety. Promoting digital literacy within a legal framework specifically tailored to online child protection should remain a priority, requiring sustained effort and institutionalization.

Tanzania stands to benefit greatly from cross-sectoral cooperation and international partnerships. By learning from the experiences of countries like China and Japan, Tanzania can enact child-specific online protection laws and strengthen its response strategies. This will ensure that children across the nation can navigate the digital world safely. Future research should focus on innovative methods to protect children online, enabling them to thrive in the digital era.






% \end{thebibliography}
\end{document}